\documentclass[]{scrartcl}
\usepackage{graphicx}
\usepackage{subfig}
\usepackage{hyperref}

\title{A note on the magnetic spatial forcing of a ferrofluid layer}

\author{Th. Friedrich\thanks{Experimentalphysik V, Universit\"at Bayreuth, D-95440, Germany}, A. Lange\thanks{Lehrstuhl Magnetofluiddynamik, TU Dresden, D-01062, Germany}, I. Rehberg\footnotemark[1], R. Richter\footnotemark[1]}

\publishers{(Submitted to Magnetohydrodynamics)}

\begin{document}
\maketitle


\begin{abstract}

We report on the response of a thin layer of ferrofluid to a spatially modulated magnetic field. This field is generated by means of a constant current in a special arrangement of aluminum wires. The full surface profile of the liquid layer is recorded by means of the absorption of X-rays. The outcome is analyzed particularly with regard to the magnetic self focusing effect under a deformable fluid layer.

\end{abstract}


\section*{Introduction}

A beautiful example of spatial forcing in a pattern forming system was studied experimentally in electroconvection\,\cite{lowe1983}. More recently, inclined layer convection was measured under the influence of lamellar surface corrugations \cite{seiden2008}. In both cases, stripes are the first convection pattern beyond a threshold.
The Rosensweig instability in a layer of ferrofluid can provide a primary instability to hexagons if a homogeneous magnetic field normal to the flat surface is applied\,\cite{cowley1967,rosensweig1985}. In case of a tilted magnetic field, a primary instability to stripes can be observed\,\cite{reimann2005}.
This system allows to study the response of both configurations to a stripe like modulation of the magnetic induction.

As a start, here we present and characterize the influence of the modulation on the layer of ferrofluid in the subcritical regime.
One method of generating a spatially modulated magnetic field was described in Refs.~\cite{beetz2008,friedrich2010} and uses rods of metallic iron to modulate a homogeneous magnetic field.
It lacks the possibility to control the amplitude of modulation independently from the field offset, a nuisance which is overcome in the experiments presented in this paper.


\section{Setup}

\begin{figure}
\begin{center}
\subfloat[]{\label{pic:grid}\includegraphics[height=4.5 cm]{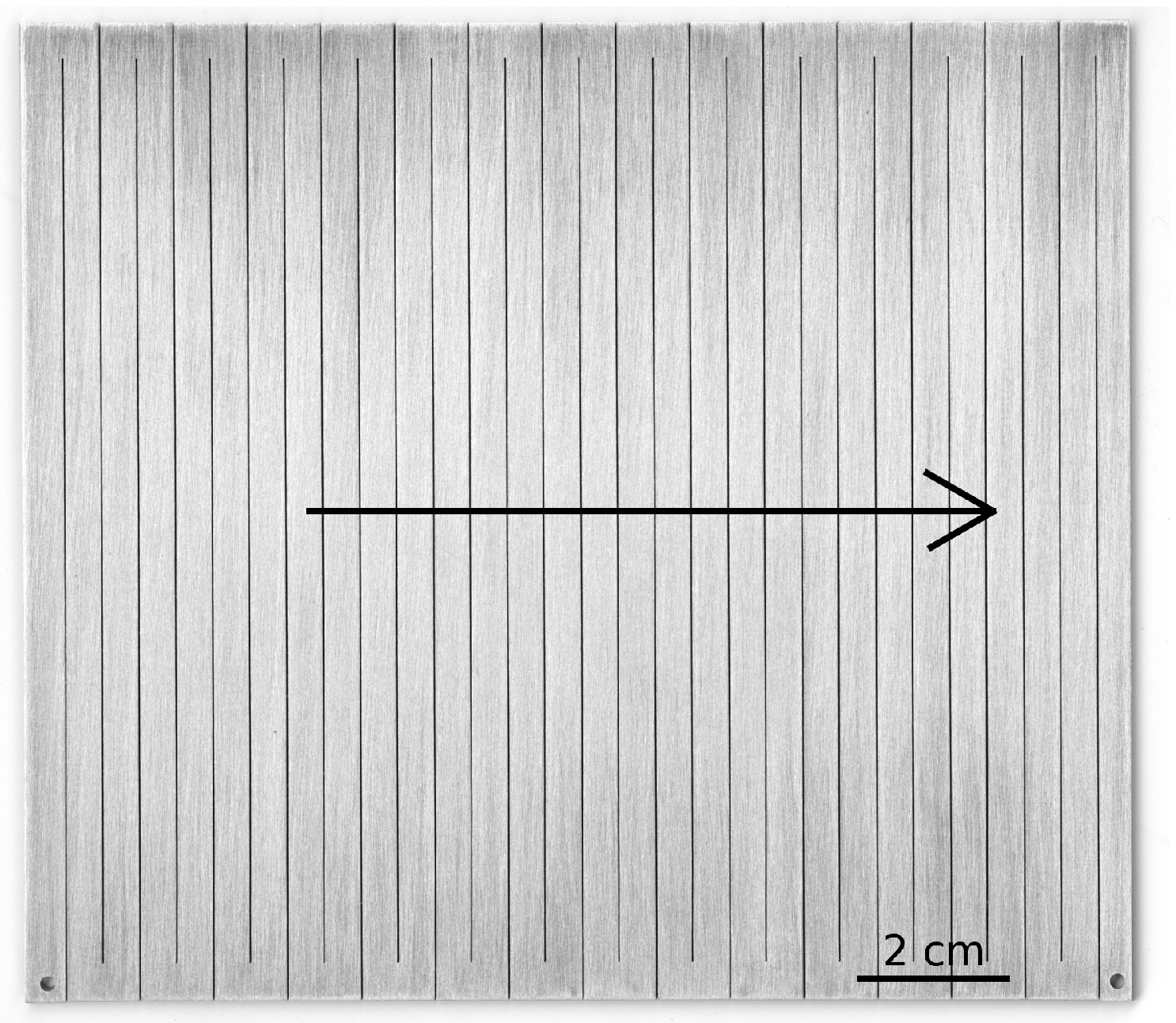}}~~~~~
\subfloat[]{\label{fig:field_graph}\includegraphics[height=4.5 cm]{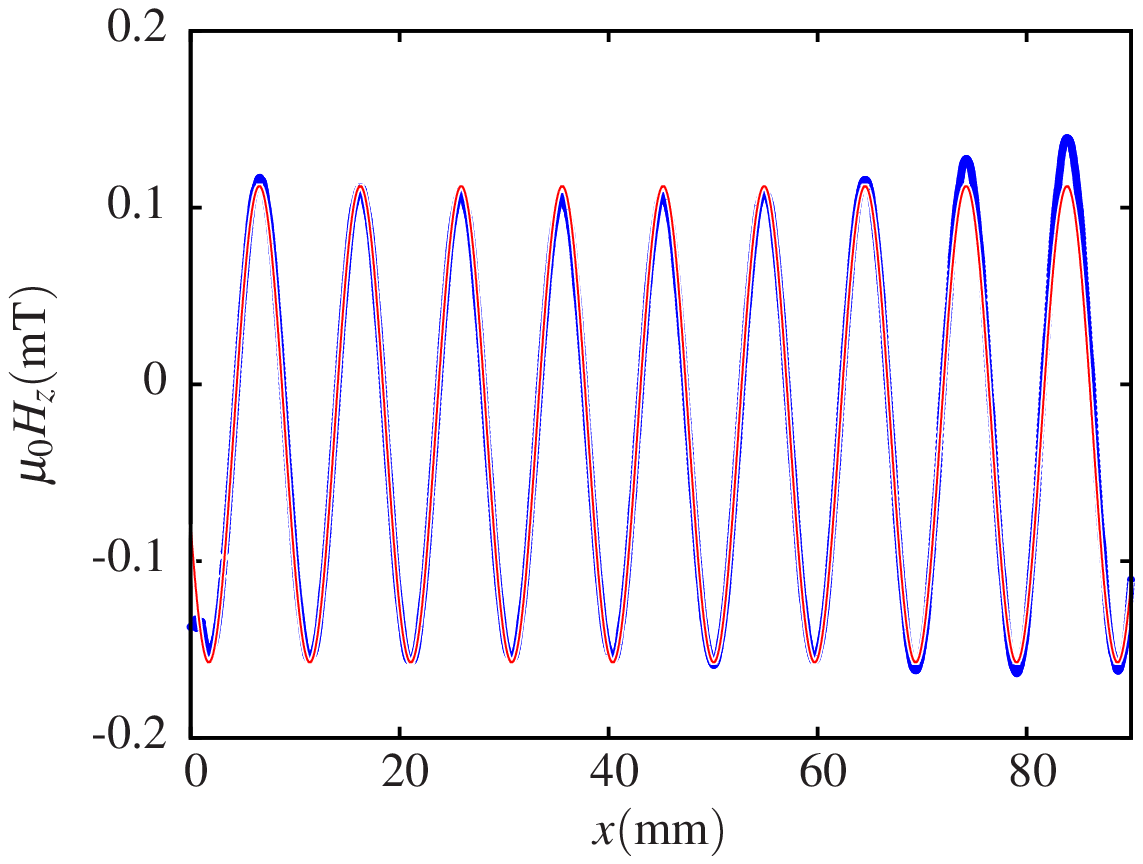}}
\end{center}
\caption{(a) Photograph of the field modulation array made of aluminum. Length: $14.6\,\mathrm{cm}$, width: $13\,\mathrm{cm}$, thickness: $2\,\mathrm{mm}$. (b) Blue: Magnetic induction $B_z=\mu_0 H_z$ generated by the modulation array at a constant current of $10\,\mathrm{A}$, measured via a Hall probe along the black arrow in \subref{pic:grid}. The distance between the array and the Hall probe was $2\,\mathrm{mm}$. Red: $H_z$ from Eq.~(\ref{eq:field}) fitted to the experimental data.}
\label{fig:grid_field}
\end{figure}
To produce a spatially modulated magnetic field, we developed an array of nonmagnetic conducting wires shown in Fig.~\ref{fig:grid_field}\subref{pic:grid}. The array is made of a plate of aluminum by cutting slits from opposite directions using electrical discharge machining. Aluminum has been chosen because of its low mass absorption coefficient for X-rays. The rectangular cross section of the wires is crucial to guarantee a spatially homogeneous absorption. By design, the current in the aligned conductors is reversing its direction from one to the next conductor.
The resulting induction $\mu_0 H_z$ is depicted in Fig.~\ref{fig:grid_field}\subref{fig:field_graph}. It was measured using a Hall probe for a certain region of the array as indicated by the arrow in Fig.~\ref{fig:grid_field}\subref{pic:grid}. To achieve a spatial forcing with a wavelength of $\lambda_c=9.6\,\mathrm{mm}$, which corresponds to the critical wavelength\,\cite{rosensweig1985} of the used ferrofluid, the center to center distance between adjacent conductors has to be $\lambda_c / 2$.
The components of the generated magnetic field $\vec{H}$ can be well approximated by
\begin{equation}\label{eq:field}
H_x = -\Delta H\cos (kx) {\rm e}^{-k z}
\qquad \mathrm{and} \qquad
H_z =H_0 + \Delta H \sin (kx) {\rm e}^{- k z},
\end{equation}
where $\Delta H$ represents the amplitude of the field modulation, $k=2 \pi / \lambda$ is the wave number and $H_0$ states an offset.
As Fig.~\ref{fig:grid_field}\subref{fig:field_graph} shows, a fit of $H_z$ in Eq.~(\ref{eq:field}) to the experimental data reveals a good agreement.
In addition to the spatially modulated field from the array described above, we superimpose a homogeneous magnetic field $H_0$ created by a pair of Helmholtz coils. Thus, $H_0$ and $\Delta H$ can be controlled by two independent currents.

\begin{figure}[ht]
\begin{center}
\includegraphics[height=4 cm]{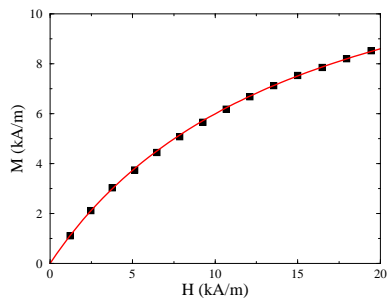}
\end{center}
\caption{Magnetization of the ferrofluid EMG909 as a function of the internal magnetic field. The points denote experimental data, and the curve is a fit of Eq.~(\ref{eq:vislovich}).}
\label{fig:mcurve}
\end{figure}

The described modulation array is mounted below the bottom of a container machined from Perspex\texttrademark. It contains a block shaped cavity with a length of $120\,\mathrm{mm}$, a width of $100\,\mathrm{mm}$ and a height of $25\,\mathrm{mm}$. This cavity is filled with $30\,\mathrm{ml}$ of ferrofluid EMG909 from Ferrotec Corporation (density $ \rho=994.5\,\mathrm{kg} / \mathrm{m^3} $, surface tension $23.37\,\mathrm{mN} / \mathrm{m}$).
The ferrofluid EMG909 consists of magnetite particles dispersed in kerosene. Its nonlinear magnetization curve is plotted in Fig.~\ref{fig:mcurve}. We measured $M(H)$ using a commercial vibrating sample magnetometer (LakeShore VSM 7404). To represent $M(H)$ for modeling, we used the approximation by Vislovich et al. for a nonlinear magnetization \cite{vislovich1990}. It describes the magnetization as
\begin{equation}
 M_{\mathrm{vis}}(H)=M_{\mathrm{sat}} \frac{H}{H+\frac{M_{\mathrm{sat}}}{\chi_{\mathrm{i}}}} \; .
\label{eq:vislovich}
\end{equation}
The fit yields the two parameters $M_{\mathrm{sat}}=15.2\,\frac{\mathrm{kA}}{\mathrm{m}}$ and $\chi_{\mathrm{i}}=0.99$.


\section{Experimental Results}

\begin{figure}
\begin{center}
\subfloat[]{\label{fig:surface_profile_2d}\includegraphics[height=4.0 cm]{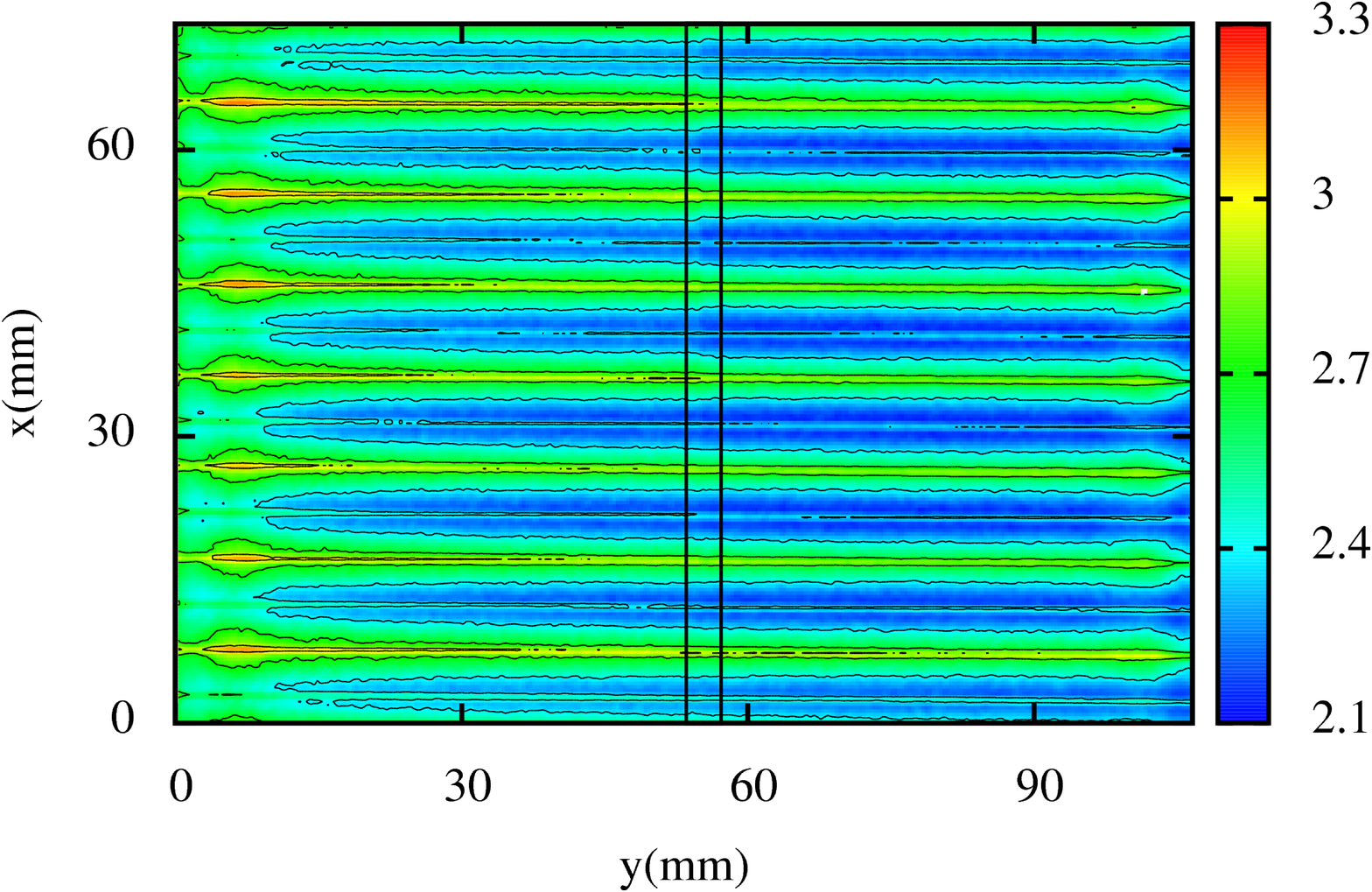}}
\subfloat[]{\label{fig:surface_profile_1d}\includegraphics[height=4.0 cm]{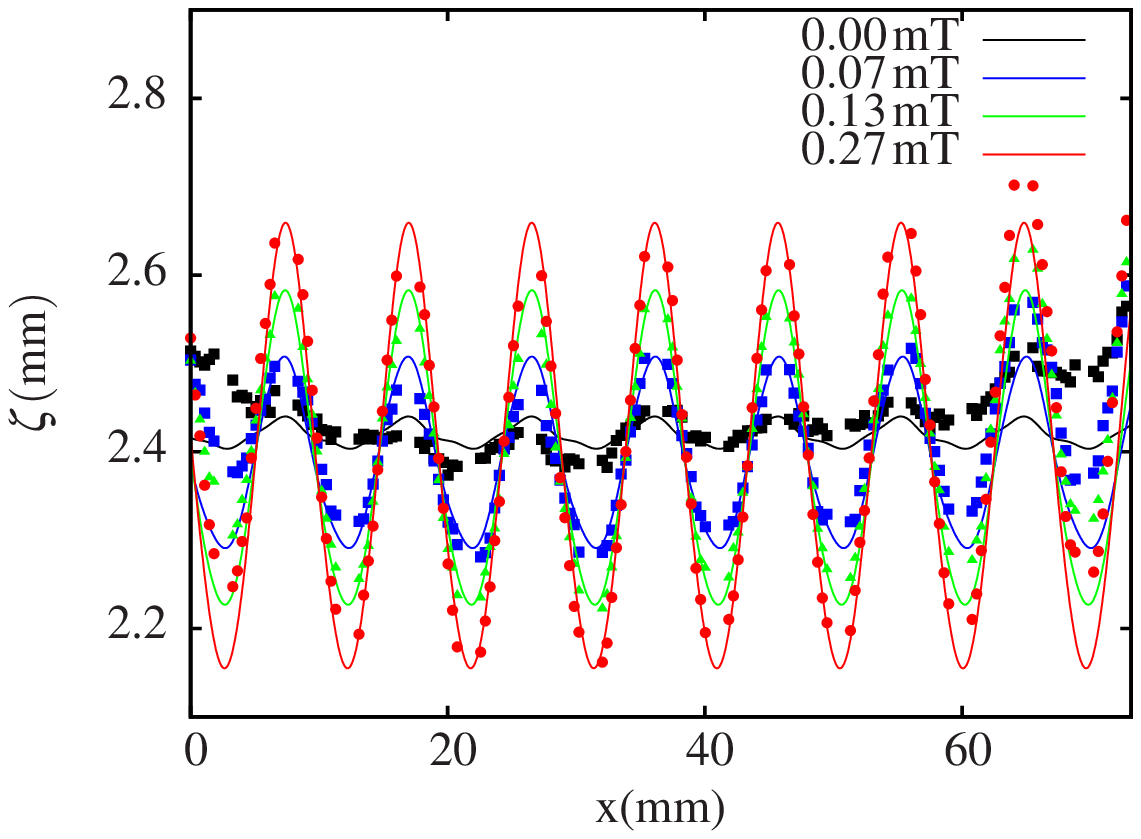}}
\end{center}
\caption{\subref{fig:surface_profile_2d} Measured height profile of the fluid layer under spatial forcing. The field offset is $\mu_0 H_0= 20.8\,\mathrm{mT}$, while the amplitude of modulation is $\mu_0 \Delta H = 0.27\,\mathrm{mT}$. The layer thickness of the liquid is color coded as shown by the indicator bar. The drawn isolines are equally spaced each $0.3\,\mathrm{mm}$. The two vertical black lines mark the region displayed in \subref{fig:surface_profile_1d}.
\subref{fig:surface_profile_1d} Averaged fluid height of the innermost ten columns ($3.6\,\mathrm{mm}$) for four different values of the modulation amplitude as nominated by the inset.}
\label{fig:profiles}
\end{figure}
Applying the modulated magnetic field to the layer of ferrofluid causes the flat surface of the liquid to deform periodically. This deformation is driven by the Kelvin force since the magnetized ferrofluid is subjected to a periodically varying field gradient. To characterize the deformation, we record X-ray transmission images of the fluid layer to measure its thickness for the whole surface at the same time. This method has been described in detail in Refs.~\cite{richter2001,gollwitzer2007}.
The deformation amplitude of the ferrofluid's free surface depends on $\Delta H$ and $H_0$. To resolve this dependance, the field offset is kept constant at $\mu_0 H_0= 20.8\,\mathrm{mT}$, while the modulation amplitude $\Delta H$ is successively increased from zero. As the amplitude of the field modulation increases, liquid ridges build up. Figure~\ref{fig:profiles}\subref{fig:surface_profile_2d} depicts the deformed surface for a modulation of $\mu_0 \Delta H = 0.27\,\mathrm{mT}$.

As the black symbols in Fig.~\ref{fig:profiles}\subref{fig:surface_profile_1d} indicate, the surface already shows some periodic deformation in the absence of a spatial modulation of the applied magnetic field. In order to measure the amplitude of the deformation, the data are fitted by a phase shifted sine with two higher harmonics (wave numbers $k$, $2k$ and $3k$) and a constant offset.
As a measure for the amplitude of deformation, only the amplitude of the fundamental mode $k$ is taken into account. The resulting dataset is depicted in Fig.~\ref{fig:amplitude.stripes}.


\section{The self focusing effect}

Under the deformed surface of a polarizable fluid, a self focusing of the magnetic field occurs. To characterize this effect with a convenient parameter, we first calculate the field $H_{\mathrm{flat}}$ under the artificial assumption of a flat surface. The parameter is then defined as the dimensionless ratio between the realistic field and $H_{\mathrm{flat}}$.

\begin{figure}[ht]
\begin{center}
\subfloat[]{\label{fig:amplitude.stripes}\includegraphics[height=4.5 cm]{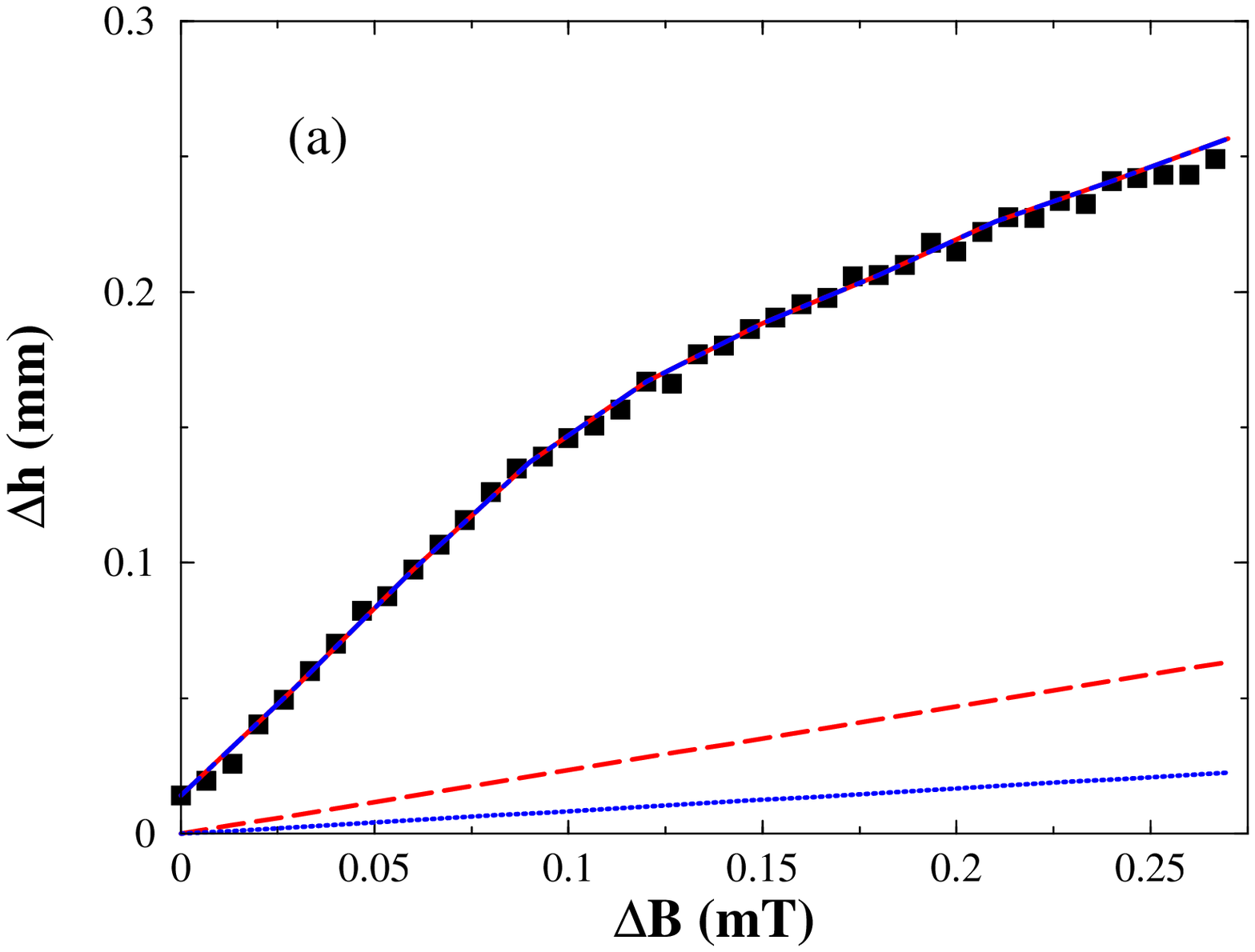}}
\subfloat[]{\label{fig:field_concentration}\includegraphics[height=4.5 cm]{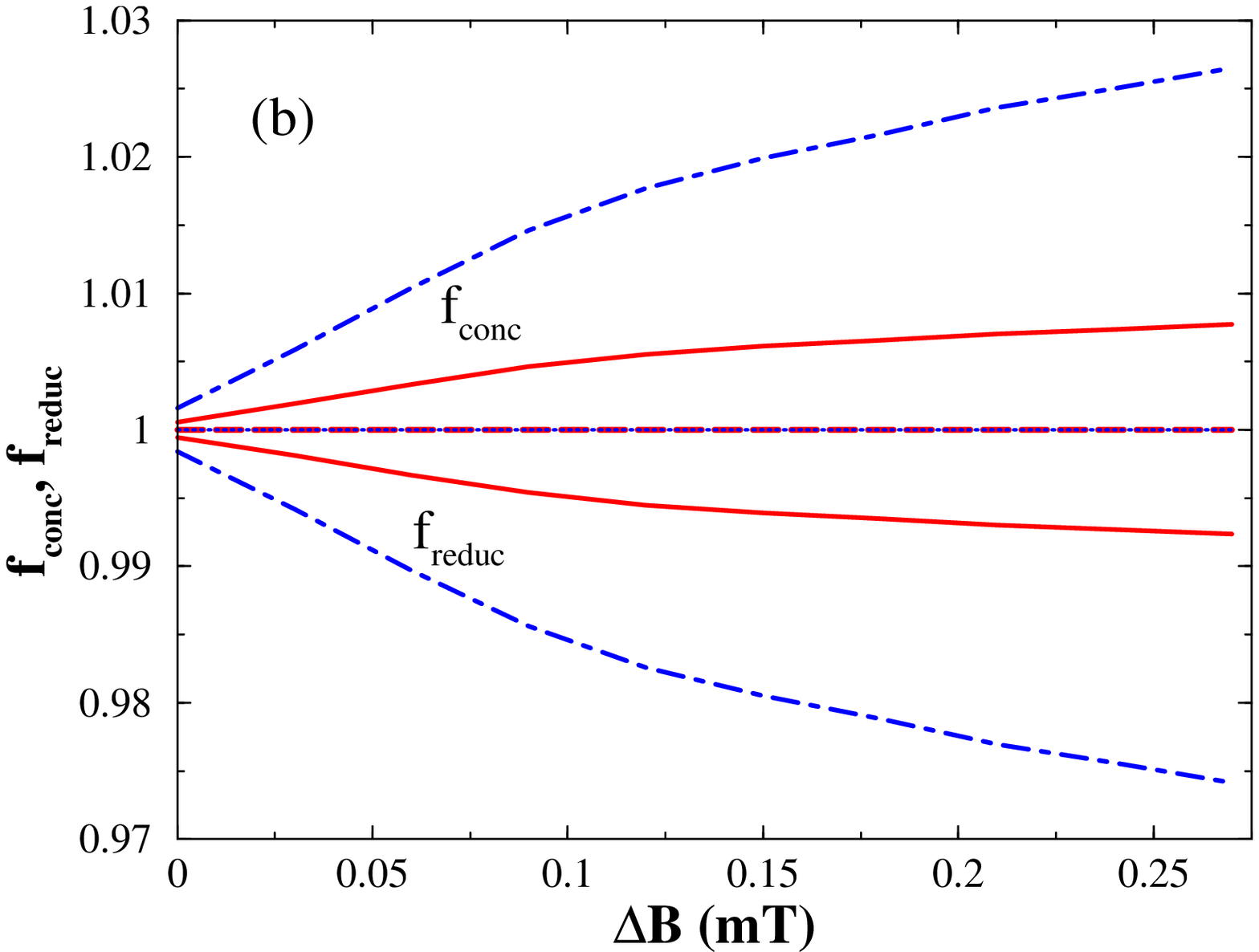}}
\end{center}
\caption{\subref{fig:amplitude.stripes} Solid squares: Measured amplitude of the surface deformation for $\Delta B=\mu_0 \Delta H$ (measured in absence of the ferrofluid) and $\mu_0 H_0 =20.8\,\mathrm{mT}$.
The red dashed (blue dotted) line shows the calculated amplitudes without focusing for a linear (nonlinear) magnetization law. The overlapping red solid and blue dot-dashed lines underneath the data points are calculated by using the values of $f$ shown in (b) for a linear magnetization and nonlinear magnetization, respectively.
\subref{fig:field_concentration} Values of $f$ needed to fit the measured amplitudes in \subref{fig:amplitude.stripes} for linear (red solid lines) and nonlinear (blue dash-dotted lines) magnetization. The red dashed and blue dotted lines at $f_{\rm conc}=f_{\rm reduc}=1$ correspond to the case where the focusing by a deformed surface was not taken into account.}
\label{fig:amplitudes}
\end{figure}

To derive a relation between the amplitude of the surface deformation and the modulation of the realistic field
we use the static form of the ferrohydrodynamic Bernoulli equation. The free surface must then be in an equilibrium of pressures and the thickness of the ferrofluid layer is given by $\zeta(x)$. Thus, the Bernoulli equation reads
\begin{equation}\label{eq:bernoulli}
C = \rho g \zeta(x) -\mu_0\int_0^{H(\zeta)} M dH' +\sigma K -\frac{\mu_0}{2}\left( \vec M \vec n \right)^2 \; .
\end{equation}
Here, $C$ is a constant, $g$ represents the gravitational acceleration and $\sigma$ is the surface tension. The normal vector to the surface and the associated curvature read
\begin{equation}\label{eq:normalvektor_curvature}
\vec n = \frac{1}{\sqrt{1+ \left( \partial_x \zeta \right )^2 }} \left( \begin{array}{c} -\partial_x \zeta \\ 1 \end{array} \right)
\qquad \mathrm{and} \qquad
K = \rm{div} \left(\vec n \right) \; .
\end{equation}
Equation~(\ref{eq:bernoulli}) has to be fulfilled for all points along the surface. In approximation, the surface shape shall be given by
\begin{equation}\label{eq:model_surface}
\zeta(x) = h + \Delta h\sin (k x) \; ,
\end{equation}
where $h$ denotes the thickness of the undisturbed fluid layer, and $\Delta h$ its amplitude of modulation. Now $\Delta h$ can be numerically determined as a function of $\Delta H$. The results are depicted in Fig.~\ref{fig:amplitudes}\subref{fig:amplitude.stripes}. The dashed red line gives the result for a linear magnetization law of the form $M=\chi H$, the dotted blue line shows the same for the nonlinear Vislovich approximation given by Eq.~(\ref{eq:vislovich}). Clearly this does not describe the observed amplitudes as the deformation of the magnetic liquid's surface causes a focusing of the magnetic field towards the maxima of the fluid height and accordingly a decrease of the field strength at the minima
\begin{equation}
f_{\mathrm{conc}}=\frac{H_{\mathrm{real}}(\zeta_{\mathrm{max}})}{H_{\mathrm{flat}}(\zeta_{\mathrm{max}})},
 \qquad
f_{\mathrm{reduc}}=f_{\mathrm{conc}}^{-1}\; .
\end{equation}
Here, $H_{\mathrm{real}}$ denotes the field strength obtained from the pressure equilibrium in order to get to the experimentally measured deformation amplitude.
To make a proper model, this focusing has to be taken into account. Without solving the magnetostatic problem together with the hydrodynamics, we can estimate the influence of the deformed surface on the magnetic field. Therefore we calculate the dimensionless parameters $f_{\mathrm{conc}}$ and $f_{\mathrm{reduc}}$ from the experimental data via the pressure equilibrium.
The outcome is shown in Fig.~\ref{fig:amplitudes}\subref{fig:field_concentration}.

The focusing of the magnetic field has earlier been investigated for fully grown Rosensweig cusps in Ref.~\cite{matthies2002}. There, a factor of $1.5$ in the alteration of the field has been found. As our deformations are about one order of magnitude smaller, a change in the magnetic field of about 3\% seems realistic.

Experimentally, we observe a deformation of the surface even without any field modulation. This can be understood as a spatial oscillatory decay induced by the meniscus at the walls of the fluid container in the advent of the Rosensweig instability. Such an effect has been measured previously \cite{richter2001,gollwitzer2007}. The wave number $k_c$ is favored by the nonmonotonous dispersion relation for surface waves on a ferrofluid in a magnetic field. Note, that the width of the container is close to ten times the critical wavelength ($\lambda_c=9.6\,\mathrm{mm}$) of the used ferrofluid.

\section{Conclusion and Outlook}
The described experiment could successfully produce a well controlled spatially modulated magnetic field and allowed to study the response of a layer of ferrofluid to this field. Using radioscopy, accurate data about the deformed surface could be gathered. The measured amplitude of the resulting deformation serves to estimate the amount of field focusing. This gives a conception of the impact of this effect, as in this case the focusing amplifies the field modulation by almost one order of magnitude compared to the case of a flat fluid layer. It remains to be investigated how this amplification depends on $H_0$. Since the presented calculation uses experimental data to determine the influence of the deformed surface on the magnetic field further studies are necessary in order to calculate the equilibrium shape of the fluid ab initio.
The modulation device described in this paper is ideally suited to force Rosensweig patterns which will be exploited in forthcoming investigations.

\section*{Acknowledgements}
We are grateful to DFG FOR 608 for financial support, Ch.\,Kruelle for inspiring discussions and K.\,Oetter for making the field modulation device possible.


\end{document}